\documentclass[10pt,conference]{IEEEtran}

\usepackage{amsfonts}
\usepackage{amssymb}
\usepackage{mathrsfs}
\usepackage{verbatim} \usepackage{times}
\usepackage[centertags]{amsmath}
\usepackage{graphicx} 

\newtheorem{thm}{\bf Theorem}
\newtheorem{cons}{Construction}[section]

\newtheorem{defn}{{Definition}}

\newtheorem{eg}{Example}[section] 

\begin{document}

\title{Signal Set Design for Full-Diversity Low-Decoding-Complexity Differential Scaled-Unitary STBCs}

\author{
\authorblockN{G. Susinder Rajan}
\authorblockA{ECE Department \\
Indian Institute of Science \\
Bangalore 560012, India \\
susinder@ece.iisc.ernet.in}
\and
\authorblockN{B. Sundar Rajan}
\authorblockA{ECE Department \\
Indian Institute of Science \\
Bangalore 560012, India \\
bsrajan@ece.iisc.ernet.in}
}
\maketitle
\begin{abstract}
The problem of designing high rate, full diversity noncoherent space-time block codes (STBCs) with low encoding and decoding complexity is addressed. First, the notion of $g$-group encodable and $g$-group decodable linear STBCs is introduced. Then for a known class of rate-1 linear designs, an explicit construction of fully-diverse signal sets that lead to four-group encodable and four-group decodable differential scaled unitary STBCs for any power of two number of antennas is provided. Previous works on differential STBCs either sacrifice decoding complexity for higher rate or sacrifice rate for lower decoding complexity.
\end{abstract}

\section{Introduction}
\label{sec1}

\PARstart{I}{t} is well known that multiple antenna systems can offer increased data rate and reliability as compared to single antenna systems when the fading coefficients are known at the receiver. However, in practice, learning the fading coefficients becomes increasingly difficult as either the fading rate or number of transmit antennas increases. Motivated by this problem, in \cite{HoS,Hug}, a transmission strategy called differential unitary space-time modulation was introduced for the noncoherent MIMO channel where neither the transmitter nor the receiver has knowledge of the channel. Essentially, using this strategy the problem of noncoherent space-time coding becomes similar to the problem of coherent space-time coding with the additional requirement for unitary codewords. Since the introduction of differential space-time codes, several works including \cite{Ogg}-\cite{YGT2} and the references in them  have focused along different directions to obtain full diversity differential space-time codes (DSTCs). Most of these previous works obtained full diversity DSTCs by neglecting the issue of encoding and decoding complexity which are crucial for practically realizing high rate systems. Though few works \cite{JaT}-\cite{YGT2} have addressed this issue partially, there seems to be no systematic  construction of high rate full diversity DSTCs guided by the requirement for low encoding and decoding complexity. 

The differential encoding/decoding setup utilized in \cite{TaC,YGT2} is more general than the differential unitary space-time modulation scheme originally proposed in \cite{HoS,Hug} in the sense that those originally proposed demand all the codeword matrices to be unitary whereas the generalized one asks for only scaled unitary codeword matrices. In this paper, we design signal sets for the rate-1 linear designs proposed in \cite{RTR} thus leading to four-group decodable differential scaled-unitary STBCs with full-diversity. 

The main contributions  can be summarized as follows:
\begin{itemize}
\item The notion of $g$-group encodable linear space-time codes is formally introduced and the inter-relationship with $g$-group decodable linear space-time codes is made clear.
\item Explicit construction of fully diverse signal sets leading to scaled-unitary codewords is provided for the designs in \cite{RTR} for arbitrary transmission rate and dimensions being a power of two. Previous algebraic approaches \cite{Ogg,OgH1} involved intensive computations which was code specific and did not permit an explicit closed form solution for arbitrary rate and dimension.
\item  The resulting codes trade off rate and decoding complexity without sacrificing either of them completely. Previous works either sacrifice decoding complexity for higher rate or sacrifice rate for lower decoding complexity.
\end{itemize}

The rest of the paper is organized as follows:  Section \ref{sec2} introduces the notion of $g$-group encodable and $g$-group decodable linear STBCs and describes its application and significance in the differential encoding/decoding setup. In Section \ref{sec3}, the  rate one complex symbols per channel use, $4$-group decodable design of \cite{RTR} is briefly described. The issues involved in the construction of fully diverse signal sets for these designs so that they are usable as differential scaled-unitary STBCs with full-diversity  are  highlighted in Section \ref{sec4} and one particular class of fully diverse signal sets is explicitly constructed for arbitrary transmission rate in bits/sec/Hz.  Section \ref{sec5} contains some concluding remarks.

\section{Preliminaries}
\label{sec2}
We first introduce the notion of $g$-group encodable and $g$-group decodable linear STBCs and explain their significance in the context of differential STBCs.

\begin{defn}
A linear design $S(s_1,s_2,\dots,s_K)$ in $K$ real indeterminates or variables $s_1,s_2,\dots,s_K$ is a $n\times n$ matrix with entries being a complex linear combination of the variables. It can be written as $S(s_1,s_2,\dots,s_K)=\sum_{i=1}^{K}s_iA_i$ where, $A_i\in \mathbb C^{n\times n}$ are called the \textit{weight matrices}. A linear STBC $\mathscr{C}$ is a finite set of $n\times n$ complex matrices which can be obtained by taking a linear design $S(x_1,x_2,\dots,x_K)$ and specifying a signal set $\mathscr{A}\subset\mathbb{R}^{K}$ from which the information vector $X=\left[\begin{array}{cccc}s_1 & s_2 & \dots & s_K\end{array}\right]^T$ take values from, with the additional condition that $S(a)\neq S(a'), \forall\ a\neq a'\in\mathscr{A}$. A linear STBC $\mathscr{C}=\left\{S(X)|X\in \mathscr{A}\right\}$ is said to be $g$-group encodable (or $\frac{K}{g}$ real symbol encodable or $\frac{K}{2g}$ complex symbol encodable) if $g$ divides $K$ and if $\mathscr{A}=\mathscr{A}_1\times\mathscr{A}_2\times\dots\times\mathscr{A}_g$ where each $\mathscr{A}_i,i=1,\dots,g\subset\mathbb{R}^{\frac{K}{g}}$.
\end{defn}
\begin{eg}
The popular Alamouti design along with square QAM constellation for each complex symbol is a $4$-group encodable linear STBC, since square QAM constellation can be realized as a Cartesian product of two PAM constellations.
\end{eg}

\subsection{Differential encoding/decoding setup}
Consider a MIMO channel with $N_T$ transmit antennas and $N_R$ receive antennas. Let $H_t$ denote the $N_T\times N_R$ channel matrix at time frame\footnote{Here the term time frame is used to denote $N_T$ channel uses.} $t$. Let $X_t$ be the transmitted $N_T\times N_T$ matrix at time frame $t$. Then the received matrix at time frame $t$ is $R_t=X_tH_t+W_t$ where, $W_t$ is the additive white Gaussian noise at the receiver at time frame $t$. The differential encoding is performed as follows. A known unitary codeword $X_0$ is first transmitted to start with. The transmitted matrix at time frame $t$ is then  $X_t=\frac{1}{a_{t-1}}U_tX_{t-1}$
where, $U_t\in\mathscr{C}$ is the codeword containing the information at time frame $t$ which satisfies $U_t^HU_t=a_t^2I$. In other words, we restrict the code $\mathscr{C}$ to contain only scaled unitary matrices. Note that the differential STBC schemes in \cite{HoS,Hug} further restrict all the codewords to be unitary matrices to ensure that the power does not tend to zero or infinity. However, even if we allow scaled unitary codewords it is possible to ensure that the average transmit power constraint say $P$ is met by requiring that $\mathrm{E}(X_t^HX_t)=\mathrm{E}(U_t^HU_t)=\mathrm{E}(a_t^2)=P.$

For such systems, a near-optimal differential decoder has been utilized in \cite{TaC,YGT2} 
which detects $U_t$ as follows:
\begin{equation}
\label{eqn_decoder}
\hat{U_t}=\mathrm{arg} \min_{U_t\in\mathscr{C}}\parallel R_t-\frac{1}{a_{t-1}}U_tR_{t-1}\parallel^2
\end{equation}
\noindent
where, $a_{t-1}$ can be estimated from the previous decision $\hat{U}_{t-1}.$ Note that the channel matrix $H$ is not required for decoding $U_t$. Further, it has been shown \cite{TaC,YGT2} that the code design criteria for full diversity and coding gain is same as in the case of unitary differential STBCs, i.e, the well known rank and determinant criteria. Also note that in general $|\mathscr{C}|$ computations are required to perform the decoding.

To reduce the encoding complexity our strategy would be to choose $\mathscr{C}$ to be a linear STBC. Let $\mathscr{C}=\left\{S(X)|X\in \mathscr{A}\right\}$. Now the higher the value of $g$, the lower the encoding complexity. Moreover, decoding $U_t$ is same as decoding the information symbol vector $X=\left[\begin{array}{cccc}s_1 & s_2 & \dots & s_K\end{array}\right]^T$. Towards obtaining the conditions for low decoding complexity, we shall first briefly introduce the notion of $g$-group decodable linear STBCs \cite{KaR2}. Though, $g$-group decodable STBCs have been studied in previous works \cite{KaR2}, the strong inter-relationship between encoding complexity and decoding complexity was not highlighted and it was implicitly assumed. Further the notion of encoding complexity was not put in formal terms.

\subsubsection{$g$-group decodable linear STBCs}
Suppose we partition the set of weight matrices of $S(X)$ into $g$-groups, the $k$-th group containing $K/g$ matrices and also the information symbol vector as, $X=\left[\begin{array}{cccc}X_1^T X_2^T \dots X_g^T\end{array}\right]^T$ where, $X_k=\left[\begin{array}{cccc}s_{\frac{(k-1)K}{g}+1} & s_{\frac{(k-1)K}{g}+2} & \dots & s_{\frac{kK}{g}}\end{array}\right]^T$, then $S(X)$ can be written as,
$$
S(X)=\sum_{k=1}^{g}S_k(X_k),\quad S_k(X_k)=\sum_{i=\frac{(k-1)K}{g}+1}^{\frac{kK}{g}}s_iA_i.
$$

Minimizing
\begin{equation}
\label{eqn_metric}
\parallel R_t-\frac{1}{a_{t-1}}S(X)R_{t-1}\parallel^2
\end{equation}
is in general not same as minimizing
\begin{equation}
\label{eqn_submetric}
\parallel R_t-\frac{1}{a_{t-1}}S_k(X_k)R_{t-1}\parallel^2
\end{equation}
for each $1\leq k\leq g$ individually. However if it so happens, then the decoding complexity is reduced by a large amount. Note that it is not possible to compute \eqref{eqn_submetric} unless the code is $g$-group encodable also.

\begin{defn}
A linear STBC $\mathscr{C}=\left\{S(X)|X\in \mathscr{A}\right\}$ is said to be $g$-group decodable (or $\frac{K}{g}$ real symbol decodable or $\frac{K}{2g}$ complex symbol decodable) if it is $g$-group encodable and if its decoding metric in \eqref{eqn_metric} can be simplified as in \eqref{eqn_submetric}.
\end{defn}
\begin{thm}
A linear STBC $\mathscr{C}=\left\{S(X)|X\in \mathscr{A}\right\}$ is $g$-group decodable if the following two conditions are satisfied.
\begin{enumerate}
\item $\mathscr{C}$ is $g$-group encodable
\item If $A_i$ and $A_j$ are the weight matrices of two variables belonging to two different groups, then they should satisfy the following equation $A_i^HA_j+A_j^HA_i=0.$
\end{enumerate}
\end{thm}
\begin{proof}
Proof is straightforward and identical to the proof in \cite{KhR1}.
\end{proof}

In the light of the definition of encoding complexity, for the Cayley codes \cite{HaH2,OgH1} if we look at the matrices obtained after applying Cayley transform, the encoding complexity is exponential. In this paper, we have taken the viewpoint of defining encoding complexity of the matrices which are used to perform differential encoding. Moreover, Cayley transform requires appropriate computation of matrix inverses.

\subsection{Problem Statement}
The differential STBC design problem is to design a linear STBC $\mathscr{C}=\left\{S(X)|X\in \mathscr{A}\right\}$ such that
\begin{enumerate}
\item All codewords are scaled unitary matrices and the average scale factor should meet the power constraint.
\item $K$ and $g$ are maximized
\item $\min_{S_1,S_2\in\mathscr{C}}|S_1-S_2|$ is maximized.
\end{enumerate}
We now briefly highlight the various issues involved in satisfying the above stated requirements by illustrating with some examples.
\begin{eg}
Let us consider the Golden code for $2$ transmit antennas. It has $8$ real variables. For the coherent MIMO channel, the signal set used is QAM for each complex variable. Hence this code is a $8$-group encodable (since QAM is a Cartesian product of two PAM signal sets) and $1$-group decodable linear STBC and thus has low encoding complexity. However, if we now impose the requirement for scaled unitary codewords, then we will have to solve for signal sets which will yield scaled unitary codewords inside the division algebra. Although this approach can potentially offer excellent coding gain, it may amount to entangling all the $8$-real variables which will make the code $1$-group encodable and $1$-group decodable. This approach was recently attempted in \cite{Ogg}.
\end{eg}
\begin{eg}
\label{eg_alamouti}
Let us take the example of the Alamouti code for $2$ transmit antennas. It has $4$ real variables. Now if we choose the signal set to be PSK (points on the unit circle) for every complex variable, then all the codewords become unitary matrices, since the Alamouti code is an orthogonal design. Hence such a code is $2$-group encodable as well as $2$-group decodable. Further this code also provides full diversity. However, note that if we take square QAM to be the signal set for each complex variable, then we get a $4$-group encodable (square QAM is a Cartesian product of two PAM signal sets) and $4$-group decodable full diversity code, but now the codewords are scaled unitary matrices as opposed to unitary matrices. Thus relaxing the codewords to be scaled unitary matrices allows us to lower the encoding and decoding complexity.
\end{eg}
The above two examples show that {\it the choice of signal sets is crucial in obtaining low encoding and decoding complexity}. 
\section{A $4$-group decodable design}
\label{sec3}
In this section, we briefly describe the construction of a rate-one linear which satisfies the conditions for $4$-group decadability. This construction was first proposed in \cite{RTR}.

Given a $n\times n$ linear design $A(x_1,x_2,\dots,x_K)$ in $K$ complex variables $x_1,x_2,\dots,x_K$, one can construct a new $2n\times 2n$ linear design $D$ as follows.
$$
\left[\begin{array}{ll}A(x_1,x_2,\dots,x_K) & B(x_{K+1},x_{K+2},\dots,x_{2K})\\B(x_{K+1},x_{K+2},\dots,x_{2K}) & A(x_1,x_2,\dots,x_K) \end{array}\right]
$$
\noindent
where, the linear design $B(x_{K+1},x_{K+2},\dots,x_{2K})$ is identical to the linear design $A(x_1,x_2,\dots,x_K)$ except that it is in different variables $x_{K+1},x_{K+2},\dots,x_{2K}$. We call this construction as the 'ABBA construction'. This construction was first introduced in \cite{TBH}, albeit only for Alamouti design.

Given a $n\times n$ linear design $A(x_1,x_2,\dots,x_K)$ in $K$ complex variables $x_1,x_2,\dots,x_K$, one can also construct a new $2n\times 2n$ linear design $S$ as follows.
$$
\left[\begin{array}{ll}A(x_1,x_2,\dots,x_K) & -B^H(x_{K+1},x_{K+2},\dots,x_{2K})\\B(x_{K+1},x_{K+2},\dots,x_{2K}) & A^H(x_1,x_2,\dots,x_K) \end{array}\right]
$$
\noindent
where, the linear design $B(x_{K+1},x_{K+2},\dots,x_{2K})$ is identical to the linear design $A(x_1,x_2,\dots,x_K)$ except that it is in different complex variables $x_{K+1},x_{K+2},\dots,x_{2K}$. We call this construction as the 'doubling construction'. This construction has also been reported in \cite{KiR3}.

We are now ready to describe our iterative construction. For $\lambda=1$, we have the Alamouti design.
\begin{cons}
\label{cons_design}\cite{RTR}
For $\lambda>1$, consider the linear design $C_1(x_1,x_2)=\left[\begin{array}{cc}x_1 & x_2\\x_2 & x_1\end{array}\right].$  Now, to obtain a linear design for $N_T=2^\lambda,\lambda>1$, we follow the steps given below.

Step 1: Starting with $C_1$, keep applying ABBA construction iteratively on it till a $2^{\lambda-1}\times 2^{\lambda-1}$ linear design $C$ is obtained.
Step 2: Then apply doubling construction on $C$ to obtain the required design.
\end{cons}
A detailed description of these and the proof that the designs given by the above construction are $4$-group decodable is given in \cite{RTR}.

\begin{eg}
Now, the design for $4$ transmit antennas according to Construction \ref{cons_design} is   
\begin{equation*}
S=\left[\begin{array}{ccrr}x_1 & x_2 & -x_3^* & -x_4^*\\
x_2 & x_1 & -x_4^* & -x_3^*\\ x_3 & x_4 & x_1^* & x_2^*\\ x_4 & x_3 & x_2^* & x_1^*\end{array}\right]
\end{equation*}
and the design for larger number of transmit antennas can also be easily constructed.
\end{eg}
\section{Choice of signal sets}
\label{sec4}
In this section, we construct fully diverse signal sets for the linear designs constructed in the previous subsection. The signal sets should be designed in  such a way that the following important requirements are met by the code simultaneously.
\begin{enumerate}
\item Scaled unitary codewords meeting power constraint
\item Four-group encodable and Four-group decodable
\item Difference of any two different codewords should be full rank (Full diversity)
\end{enumerate}

\begin{figure}[h]
\centering
\includegraphics[width=3in]{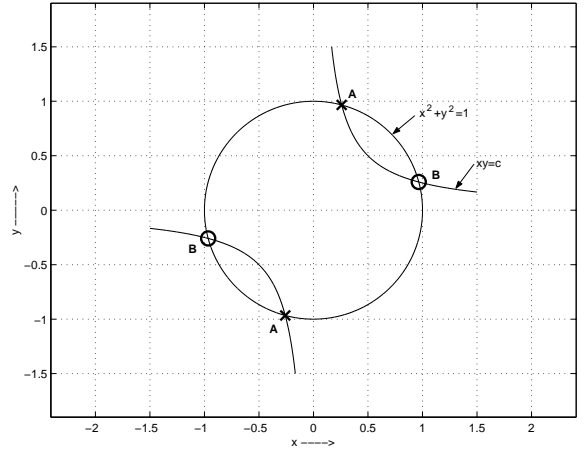}
\caption{Signal set structure in 2 dimensions}
\label{fig_signal2dimstruct}
\end{figure}

\begin{figure*}
\centering
\includegraphics{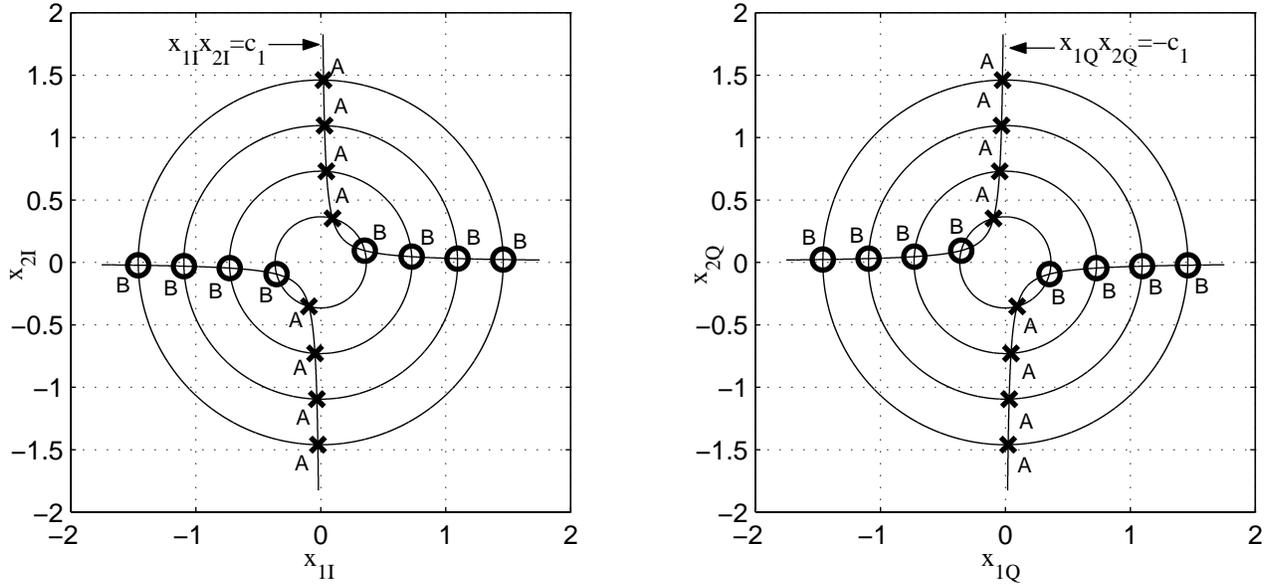}
\caption{General signal set for four transmit antennas}
\label{fig_signal2dim}
\hrule
\end{figure*}

We shall first illustrate the procedure for construction of signal sets for $4$ transmit antennas and then generalize the ideas for any $N_T=2^\lambda$ transmit antennas. For the  design for $4$ transmit antennas we study 
\begin{equation*}
S^HS=\left[\begin{array}{ccrr}
a & b & 0 & 0\\
b & a & 0 & 0\\ 
0 & 0 & a & b\\ 
0 & 0 & b & a
\end{array}\right]
\end{equation*}
where $a=\sum_{i=1}^{4}|x_i|^2$ and $b=x_1^*x_2+x_2^*x_1+x_3^*x_4+x_4^*x_3,$ to find out the conditions on the signal sets under which the codewords are scaled unitary matrices. We see that the signal set should be chosen such that the following condition is satisfied for all the signal points:
$$
x_1^*x_2+x_2^*x_1+x_3^*x_4+x_4^*x_3=0.
$$
However, we should be careful to not to disturb $4$-group encodability in the process. Hence we first identify the grouping of the variables. According to the construction of \cite{RTR}, the four groups are as follows.
$$\left\{x_{1I},x_{2I}\right\};~ \left\{x_{1Q},x_{2Q}\right\};~ \left\{x_{3I},x_{4I}\right\};~ \left\{x_{3Q},x_{4Q}\right\}$$
The chosen signal sets should be in such a way that there are no joint constraints on variables from different groups. If that happens, then the code will no longer be $4$-group encodable and $4$-group decodable. Putting together all the requirements for scaled unitary codewords, we have
$$ x_{1I}x_{2I}=-x_{1Q}x_{2Q};~~~~ x_{3I}x_{4I}=-x_{3Q}x_{4Q}.$$ 
The above equations can be satisfied without disturbing $4$-group encodability as shown below.
\begin{equation}
\label{eqn_signalset1}
\begin{array}{c}
x_{1I}x_{2I}=-x_{1Q}x_{2Q}=c_1;~ x_{3I}x_{4I}=-x_{3Q}x_{4Q}=c_2
\end{array}
\end{equation}
where, $c_1$ and $c_2$ are positive real constants. Then, the average power constraint requirement can be met by satisfying the conditions
{\small
\begin{equation}
\label{eqn_signalset2}
\mathrm{E}(x_{1I}^2+x_{2I}^2)=\mathrm{E}(x_{1Q}^2+x_{2Q}^2)=\mathrm{E}(x_{3I}^2+x_{4I}^2)=\mathrm{E}(x_{3Q}^2+x_{4Q}^2)=1 
\end{equation}
}

\noindent
where, without loss of generality\footnote{We can always scale all the constellation points according to the transmit power requirement.} we have considered the average power on a symbol to equal $1$. Solutions to \eqref{eqn_signalset1} are simply points on a hyperbola. Thus a common set of solutions of \eqref{eqn_signalset1} and \eqref{eqn_signalset2} can be obtained by taking points on the intersection of circles and hyperbolas. But we have a third requirement of full diversity which has to be met. For this we use the structure of the constructed designs. The constructed designs have the form  $S=\left[\begin{array}{cc}A & -B^H\\B & A^H\end{array}\right].$ It can be shown \cite{RTR} that $|\Delta S^H\Delta S| \geq \mathrm{max}(|\Delta A|^2,|\Delta B|^2)^2.$ Thus we can guarantee full diversity by ensuring that $\Delta x_1\neq\pm\Delta x_2$ and $\Delta x_3\neq\pm\Delta x_4$. Just like before, we should be careful not to disturb $4$-group encodability in the process. We take care of that requirement also by satisfying the following conditions:
\begin{equation}
\label{eqn_signalset3}
\begin{array}{c}
\Delta x_{1I}\neq\pm\Delta x_{2I};~ \Delta x_{1Q}\neq\pm\Delta x_{2Q};\\
\Delta x_{3I}\neq\pm\Delta x_{4I};~ \Delta x_{3Q}\neq\pm\Delta x_{4Q}.
\end{array}
\end{equation}
The solution satisfying all the three conditions \eqref{eqn_signalset1}, \eqref{eqn_signalset2} and \eqref{eqn_signalset3} can be found simply by finding the intersection of points on the unit circle $x^2+y^2=1$ with a hyperbola $xy=c$, where $c<1$ on the two dimensional $xy$ plane. This is illustrated in Fig. \ref{fig_signal2dimstruct}.
Observe that the hyperbola intersects the circle at four different points. But the full diversity criterion demands that $\Delta x\neq\pm\Delta y$. After enforcing this condition, only two points survive out of the four points. They can be either the set of points marked \textbf{A} or the set of points marked \textbf{B} in Fig. \ref{fig_signal2dimstruct}. Thus we have obtained a signal set containing $2$ points. If we need more points, we can then invoke the fact that scaled unitary codewords are sufficient. We can draw more circles (centered at origin) with radii such that the average power constraint is met and then find those points intersecting with the hyperbola. More precisely, to get $M$ points, draw $\frac{M}{2}$ concentric circles with increasing radii $r_1,r_2,\dots,r_{\frac{M}{2}}$ such that $\sum_{i=1}^{\frac{M}{2}}r_i^2=\frac{M}{2}.$  Then find those points intersecting with the hyperbola $xy=c$ where, $c$ is a positive number less than\footnote{This condition is necessary since otherwise the hyperbola will not intersect the circle with least radius.} $r_1^2$. In this manner we can get the desired signal set for the variables $x_{1I},x_{2I}$ and $x_{3I},x_{4I}$. The signal set for the variables $x_{1Q},x_{2Q}$ and $x_{3Q},x_{4Q}$ can be obtained by considering a different hyperbola. This is illustrated in Fig. \ref{fig_signal2dim}.
Now, generalizing the above ideas, it can be shown that Construction \ref{cons_signalset} gives the closed form solution of the signal sets that  satisfies all the requirements for any power of two  number of antennas.
\begin{cons}
\label{cons_signalset}
Suppose we want a $M$-points signal set $\subset\mathbb{R}^{2^{\lambda+1}}$ for the constructed design for $N_T=2^\lambda$ transmit antennas. Then, the resulting signal set $\subset\mathbb{R}^{2^{\lambda+1}}$ should be a Cartesian product of $4$ signal sets in $\mathbb{R}^{{2^{\lambda-3}}}$, since we insist on $4$-group encodability. In our case, we choose all the four sets to be identical and each contains $\sqrt[4]{M}$ points. Let the signal points in $\mathbb{R}^{{2^{\lambda-3}}}$ be labeled as $p_i,\ i=1,\dots,\sqrt[4]{M}$. If $i=2q+r$, then $p_i$ is given by
\begin{equation}
\begin{array}{c}
p_i[j]=0\ \forall j\neq(q\ \mathrm{mod}\ 2^{\lambda-3})+1\\
p_i[(q\ \mathrm{mod}\ 2^{\lambda-3})+1]=r_q,\ \mathrm{if}\ r=0\\
p_i[(q\ \mathrm{mod}\ 2^{\lambda-3})+1]=r_q,\ \mathrm{if}\ r=1
\end{array}
\end{equation}
where, for a vector $x$, $x[i]$ denotes the $i$-th entry of the vector $x$ and $r_q,\ q=1,\dots,\frac{\sqrt[4]{M}}{2}$ are positive real numbers such that $r_{q+1}>r_{q},\ \forall q=1,\dots,\frac{\sqrt[4]{M}}{2}-1$ and $\sum_{i=1}^{\frac{\sqrt[4]{M}}{2}}r_i^2=\frac{\sqrt[4]{M}}{2}$.
\end{cons}
\begin{thm}
\label{thm_signalset}
Construction \ref{cons_signalset} provides fully diverse signal sets for the designs given by Construction \ref{cons_design}.
\end{thm}
\begin{eg}
Let $N_T=2^3=8$ and $M=16^4$. Thus the rate of transmission of this code will be $\frac{\log_2 M}{8}=2$ bits per channel use. The corresponding $4$ dimensional signal set is shown below:
\begin{equation*}
\begin{array}{l}
p_1=\left[\begin{array}{cccc}r_1 & 0 & 0 & 0\end{array}\right]^T;~
p_2=\left[\begin{array}{cccc}-r_1 & 0 & 0 & 0\end{array}\right]^T\\
p_3=\left[\begin{array}{cccc}0 & r_2 & 0 & 0\end{array}\right]^T;~
p_4=\left[\begin{array}{cccc}0 & -r_2 & 0 & 0\end{array}\right]^T\\
p_5=\left[\begin{array}{cccc}0 & 0 & r_3 & 0\end{array}\right]^T;~
p_6=\left[\begin{array}{cccc}0 & 0 & -r_3 & 0\end{array}\right]^T\\
p_7=\left[\begin{array}{cccc}0 & 0 & 0 & r_4\end{array}\right]^T;~
p_8=\left[\begin{array}{cccc}0 & 0 & 0 & -r_4\end{array}\right]^T\\
p_9=\left[\begin{array}{cccc}r_5 & 0 & 0 & 0\end{array}\right]^T;~
p_{10}=\left[\begin{array}{cccc}-r_5 & 0 & 0 & 0\end{array}\right]^T\\
p_{11}=\left[\begin{array}{cccc}0 & r_6 & 0 & 0\end{array}\right]^T;~
p_{12}=\left[\begin{array}{cccc}0 & -r_6 & 0 & 0\end{array}\right]^T\\
p_{13}=\left[\begin{array}{cccc}0 & 0 & r_7 & 0\end{array}\right]^T;~
p_{14}=\left[\begin{array}{cccc}0 & 0 & -r_7 & 0\end{array}\right]^T\\
p_{15}=\left[\begin{array}{cccc}0 & 0 & 0 & r_8\end{array}\right]^T;~
p_{16}=\left[\begin{array}{cccc}0 & 0 & 0 & -r_8\end{array}\right]^T\\
\end{array}
\end{equation*}
where,
{\small
$$
\begin{array}{l}
r_1=0.3235;~
r_2=\sqrt{3}r_1;~
r_3=r_2+\frac{r_5-r_2}{3};~
r_4=r_2+2\left(\frac{r_5-r_2}{3}\right)\\
r_5=3r_1;~
r_6=\left(2+\sqrt(3)\right)r_1;~
r_7=r_3+2r_1;~
r_8=r_4+2r_1.
\end{array}
$$
}
\end{eg}

\section{Discussion}
\label{sec5}
An important direction for further research is to optimize the signal sets for maximizing the coding gain. Extending this work to general $g$-group ML decodable STBCs is also another interesting direction for further work. 

\section*{Acknowledgment}
This work was supported through grants to B.S.~Rajan; partly by the
IISc-DRDO program on Advanced Research in Mathematical Engineering, and partly
by the Council of Scientific \& Industrial Research (CSIR, India) Research
Grant (22(0365)/04/EMR-II).




\begin{thebibliography}{1}


\bibitem{HoS} B. M. Hochwald and W. Sweldens, ``Differential unitary space-time modulation," \emph{IEEE Trans. on Communications}, Vol. 48, pp. 2041-2052, Dec. 2000.


\bibitem{Hug} Brian L. Hughes, ``Differential Space Time Modulation," \emph{IEEE Trans. on Inform. Theory}, Vol. 46, No. 7, pp. 2567-2578, Nov. 2000.






\bibitem{Ogg} Fr\'{e}d\'{e}rique Oggier, ``Cyclic Algebras for Noncoherent Differential
Space-Time Coding," To appear in \emph{IEEE Trans. Inform. Theory}. Available online http://www.systems.caltech.edu/\~{}frederique/draftDiff.ps


\bibitem{HaH2} B. Hassibi and B. M. Hochwald, ``Cayley Differential Unitary Space Time Codes," \emph{IEEE Trans. Inform. Theory}, Vol. 48, No. 6, pp. 1485-1503, June 2002.

\bibitem{OgH1} Fr\'{e}d\'{e}rique Oggier and Babak Hassibi, ``Algebraic Cayley differential Space-Time Codes,'' \emph{IEEE Trans. on Inform. Theory}, Vol. 53, No. 5, May 2007. Available online http://www.systems.caltech.edu/\~{}frederique/draftcayley.ps



\bibitem{JaT} Hamid Jafarkhani and Vahid Tarokh, ``Multiple Transmit Antenna Differential Detection From Generalized Orthogonal Designs," \emph{IEEE Trans. Inform. Theory}, Vol. 47, No. 6, pp. 2626-2631, Sep. 2001.

Space Time Block Codes," \emph{IEEE Signal Processing Letters}, Vol. 9, No. 2, pp. 57-60, Feb. 2002.

\bibitem{LiX} X.-B. Liang and  X.-G. Xia , ``Fast Differential Unitary Space-Time Demodulation via Square Orthogonal Designs," \emph{IEEE Trans. on Wireless Communications}, Vol. 4, No. 4, pp. 1331-1336, July 2005.

\bibitem{TaC} M. Tao and R. S. Cheng, ``Differential space-time block codes," Proceedings of \emph{IEEE Globecom 2001}, Vol. 2, pp. 1098-1102, San Antonio, USA, 25-29, Nov. 2001.

\bibitem{ZhJ} Yun Zhu and Hamid Jafarkhani, ``Differential Modulation Based on
Quasi-Orthogonal Codes," \emph{IEEE Trans. on Wireless Communications}, Vol. 4, No. 6, pp. 3018-3030, Nov. 2005.


\bibitem{YGT2} C. Yuen; Y. L. Guan; T. T. Tjhung, ``Single-Symbol-Decodable Differential Space-Time Modulation Based on QO-STBC,'' \emph{IEEE Trans. Wireless Comms.}, Vol. 5, Dec. 2006, pp. 3329-3335.





\bibitem{KhR1} Md. Zafar Ali Khan and B. Sundar Rajan, ``Single-Symbol Maximum-Likelihood Decodable Linear STBCs," \emph{IEEE Trans. Inform. Theory}, Vol.52, No.5, pp.2062-2091, May 2006.







\bibitem{RTR} G. Susinder Rajan, Anshoo Tandon, B. Sundar Rajan, ``On Four-group ML decodable distributed space time codes for cooperative communication," Proceedings of \emph{IEEE Wireless Communications and Networking Conference (WCNC 2007)}, Hong Kong, March 11-15, 2007.

\bibitem{KaR2} Sanjay Karmakar, B.Sundar Rajan, ``Multigroup decodable STBCs from Clifford Algebras," Proceedings of \emph{IEEE International Workshop in Information Theory}, Chengdu, China, Oct.22-26, 2006, pp. 448-452.


\bibitem{KiR3} Kiran T. and B. Sundar Rajan, ``Distributed Space-Time Codes with Reduced Decoding Complexity,''Proceedings of \emph{IEEE International Symposium on Inform. Theory}, Seattle, July 9-14, 2006, pp.542-546.


\bibitem{TBH} O. Tirkkonen, A. Boariu, and A. Hottinen, ``Minimal nonorthogonality rate 1 space-time block code for 3+ Tx antennas,'' Proceedings of \emph{IEEE 6th Int. Symp. Spread-Spectrum Techniques and Applications} (ISSSTA 2000), Sept. 2000, pp. 429-432.











\end{thebibliography}
\end{document}